\documentclass[%
 aps,
preprint,
floatfix,
]{revtex4-1}

\usepackage{graphicx}
\usepackage{dcolumn}
\usepackage{bm}
\usepackage{epstopdf}
\usepackage{siunitx}
\usepackage{svg}
\usepackage{amsmath}
\usepackage{gensymb}
\usepackage{float} 

\begin{document}


\title{Geometrical design for pure current-driven domain wall nucleation and shifting}

\author{D. Osuna Ruiz *\textsuperscript{1}} 
\email{osunaruiz.david@usal.es}
\author{O. Alejos\textsuperscript{2}}
\author{V. Raposo\textsuperscript{1}}
\author{E. Martínez\textsuperscript{1}}%
 
\affiliation{%
\textsuperscript{1}Department of Applied Physics, University of Salamanca, Salamanca 37008, Spain
 }%
\affiliation{%
\textsuperscript{2}Department of Electricity and Electronics, University of Valladolid, Valladolid, Spain.
 }%
\date{\today}

\begin{abstract}
Nucleation of domain walls by current-driving a single domain wall, confined to the junction area of two symmetrical strips, is investigated using systematic micromagnetic simulations. Secondary domain walls (equivalently, bits encoded in domains) are simultaneously nucleated and driven by alternatively applying current pulses between two terminals in the structure. Simulations show that nanosecond-duration current pulses nucleate and drive series of robust \textit{up}/\textit{down} domains even under realistic conditions. These results demonstrate a technique for sequentially nucleating and shifting domain walls without using attached external \lq bit lines', fields or modifying the ferromagnetic strip.

\end{abstract}

\pacs{Valid PACS appear here}
\maketitle


A magnetic domain wall (DW) consists of a transition region of magnetization that separates two magnetic domains. These magnetic configurations are interesting, not only for their fundamental physics, but also for technological applications \cite{Parkin190,article_luo}. In particular, DW dynamics have attracted special attention in ultrathin ferromagnetic (FM) films that show high perpendicular magnetic anisotropy (PMA), sandwiched in a multilayered structure between a Heavy Metal (HM) layer and typically an oxide (Ox) layer \cite{Haazen2013DomainWD,Emori2013CurrentdrivenDO,Martnez2014CurrentdrivenDO,PMID:27882932,article_ryu}. The interfacial Dyalozinshki-Moriya Interaction (iDMI) present in those structures, promote homochiral Neel DWs separating \lq up' or \lq down' domains \cite{article_chiraldws}. These domains can be potentially exploited as information bits. An electrical current in the HM layer is used to drive the chiral Neel DWs along the FM strip via spin-orbit torques (SOTs) \cite{Emori2013CurrentdrivenDO,article_ryu}. Also, they can be compactly stored and driven along magnetic tracks \cite{article_luo}.

\par For data-recording applications, DW-based devices must be efficient in three typical stages of operation, namely, writing, moving and reading bits (domains). The second stage is typically implemented by using SOTs to shift DWs, and the third one by reading the out-of-plane magnetization using a magnetic tunnel junction (MTJ). But prior to all the above, domain and DW nucleation (or writing bits) is necessary as a first step. An Oersted magnetic field from an attached current \lq bit line' is typically used, which locally reverses magnetisation in the strip. However, the Oersted field that nucleates a new DW can also annihilate already shifted DWs \cite{doublestrip}. Note that due to the long-range dependence of the Oersted field, a standard bit-line architecture is not useful to nucleate subsequent DWs without perturbing the information already coded in up or down domain between them (see details in \cite{doublestrip}). To solve this issue, solutions based on modifying the strip geometry \cite{pi_stripline}, adding more voltage terminals \cite{cross,doi:10.1063/1.4998216_selective} or using a double \lq bit line' configuration have been proposed \cite{doublestrip}. However, these solutions still require external attachments and, a more intricate multilayer design. 

\par Other proposals for nucleation of chiral Neel DWs in PMA strips rely on electric fields or ion beam irradiation, which induces local changes in the local magnetic anisotropy \cite{ionbeam,lavrijsen}. In this way, nucleation and injection of chiral DWs have also been demonstrated by modifying the magnetic properties of the strip. For example, Dao et al. \cite{article_dao} nucleated and injected chiral DWs by a tailored \lq in/out-of-plane' magnetization boundary, strategically placed in the strip. For this, a precise local control on magnetic anisotropy is still required. Also in \cite{article_dao}, domains of $\sim$ 1 $\mu$m wide were obtained by injecting current pulses in the presence of external in-plane magnetic fields. Simpler mechanisms for nucleation of packed chiral DWs, without manipulating the magnetic anisotropy nor using magnetic fields, remains elusive \cite{packed2}, which is a key first step for developing efficient DW-based recording devices.

\par Here we explore an approach based on a single current-driven DW, confined to the junction area of two symmetrical strips, that works as a DW nucleator. The subsequent domain series is coherent with the input pulsed current without the need of external magnetic fields or attachments. In what follows, we firstly describe the micromagnetic model used for our simulations. After that, we present and discuss the results from ideal and realistic scenarios, typically found in experiments. Finally, the main conclusions are summarised.

\begin{figure}[t]
\centering 
\includegraphics[width=7.5cm]{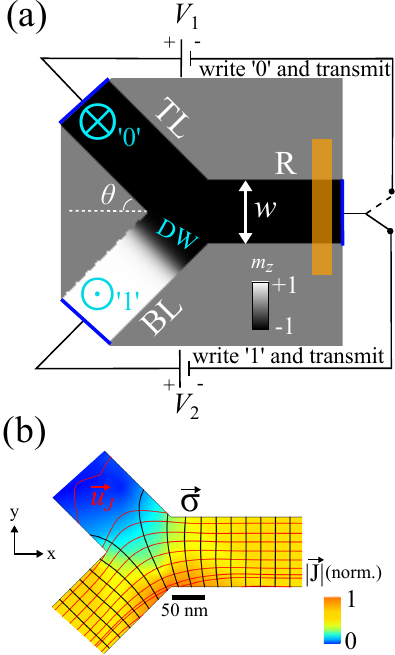}
\caption{(a) Scheme showing the symmetric strip structure modeled in simulations. The PMA configuration is shown by the \textit{up}/\textit{down} domains and the separating DW. The input current pulses, alternated between the symmetric arms, are represented by a switch and a potential difference between terminals. (b) Streamlines of $\vec{J}$ (red) and $\vec{\sigma}$ (black) are shown, obtained from COMSOL$^\text{TM}$ \cite{comsol} when $V_2$ is applied. Color scale represents the (normalized) magnitude of the resulting inhomogenous $J$.}  \label{fig1}
\end{figure}

\par Fig. 1(a) shows a schematic of the simulated HM/FM/Ox strip geometry, with width $w=96$ nm, thickness $t_{FM} = 0.6$ nm and aperture half-angle $\theta=45$º. Its initial magnetisation state at equilibrium is also shown. The geometry consists of three combined strips, two of them are symmetrical with respect the central, longitudinal strip. For convenience, hereon in we refer to them as the \lq Top-Left strip' (TL), \lq Bottom-Left strip' (BL) and \lq Register strip' (R), as depicted in Fig. 1(a). Typical Pt/Co/Ox parameters were adopted \cite{Emori2013CurrentdrivenDO,Martnez2014CurrentdrivenDO,PMID:27882932,Alejos}: exchange constant $A_{ex} = $ $1.6$ $\times$ $10^{-11}$ Jm$^{-1}$,  saturation magnetization $M_S=0.8\times 10^6$ Am$^{-1}$, anisotropy constant $K_{U} = $ $0.8$ $\times$ $10^{6}$ Jm$^{-3}$ and DMI parameter $D = $ $-1.2\times10^{-3}$ Jm$^{-2}$ \cite{article_dao}. HM layer lays below the FM layer, with the same layout. The \textit{up} and \textit{down} domains at the TL and BL strips act as \lq bit reservoirs'. A detection area (orange shaded area) at $\sim300$ nm from the junction area, represents an MTJ for reading the local out-of-plane magnetization ($m_z$). A magnetic configuration with PMA is represented, where the DW is a chiral Neel type wall \cite{article_chiraldws}. A more detailed discussion on the nucleation of the initial DW, required by standard techniques, can be found in the supplementary material.

\par Starting from the initial state shown in Fig. 1(a), current pulses ($J_1$ or $J_2$) are injected by applying voltage pulses between terminals (blue lines in Fig. 1(a)) at the TL or BL strip and the R strip (i.e. $V_1$ or $V_2$, depending on the switch position). Then, the magnetization dynamics is numerically investigated by solving the LLG equation,

\begin{gather}
\frac{d\vec{m}(t)}{dt} = -\gamma_{0}\vec{m}(t)\times \vec{H}_{eff} + \alpha \vec{m}\times \frac{d\vec{m}(t)}{dt} + \vec{\tau}_{SOT}, 
\end{gather}
where $\vec{\tau}_{SOT}$ accounts for the SOT due to the Spin Hall Effect (SHE), $\vec{\tau}_{SOT} = -\gamma_{0}H_{SL}^{0}\vec{m}\times (\vec{m} \times \vec{\sigma})$. $\vec{\sigma}$ is the spin current polarisation, that satisfies $\vec{\sigma}=\vec{u}_{z}\times\vec{u}_{J}$, where $\vec{u}_z$ is the unit vector direction along the spin current $\vec{J}_S$ and $\vec{u}_J$ is that of the applied current $\vec{J}$. $H_{SL}^{0}=\frac{\hbar \theta_{SH} J}{2|e|\mu_{0}M_{S}t_{FM}}$ defines the Slonczewski-Like-SOT magnitude, which is proportional to $J$. Micromagnetic simulations were performed using Mumax3 \cite{doi:10.1063/1.4899186}, considering SHE angle $\theta_{SH}=0.12$ and Gilbert damping constant $\alpha$ $=$ $0.1$ \cite{Alejos}. Cell sizes along $x,y$ and ${z}$ were fixed to 1 nm, 1 nm and 0.6 nm, respectively. A more detailed description of the micromagnetic model can be found elsewhere \cite{angular_dependence}. Fig. 1(b) shows streamlines for an inhomogenous $\vec{J} = \vec{J}(\vec{r})=J(x,y)\vec{u}_J$ as well as for $\vec{\sigma}$ when $V_2=1$ V and $V_1 = 0$ V are applied, obtained from COMSOL$^\text{TM}$ \cite{comsol}, for $\theta=45$º. 

\par Sequential nucleation of domains in the R (register) strip can be characterised as phase diagrams for each combination of $J$ and pulse duration $d$. Micromagnetic results reveal that two different \lq states' can be obtained. These states will determine the success of nucleating domains in sequence. Fig. 2(a)-(b) show two exemplary cases, obtained after applying the pulse sequences depicted in the top panels, for $J=2.5$ TA/m$^2$ and the indicated $d$ and $\tau$ (time between consecutive pulses) in each case. Their related output signals ($m_z$) at the reading position are shown in the middle panels. Bottom panels in Fig. 2 (a)-(b) show timestamps of $m_z$ for the respective $d$ and $\tau$ in each case. For $d=0.2$ ns (Fig. 2(a)), the DW does not reach the opposite edge of the strip, and relaxes into a new equilibrium position in the BL strip during $\tau$. For $d=0.5$ ns (Fig. 2(b)), the DW reaches the edge within the duration of each pulse, before splitting and nucleating a DW in the R strip. Only the latter situation leads to an operational DW nucleating performance where the nucleated domains are correlated with the input sequence of $J$. For simplicity, we call these two possible outcomes: \lq \textbf{U}n-triggered' state (Fig. 2(a)), and \lq \textbf{N}ucleator' state (Fig. 2(b)).

\par This purely geometrical method can generate any domain sequence, as the same process shown in Fig. 2(b) can be replicated in next pulse iterations. The working principle is as follows: a voltage pulse $V_2$ ($V_1$) is applied between terminals of the BL (TL) and R strips, generating a both \lq nucleating' and \lq driving' current $J_2$ ($J_1$). The generated current drives a DW, previously nucleated at the BL or TL strip (BL in Fig. 2), into the R strip. The nucleation of such first DW is performed only once by standard ways, for example by injecting a current along a bit line (see the additional simulations and related discussion in the supplementary material). Driven by $J$ ($J_2$ in Fig. 2), the DW will eventually stop before reaching the start of the opposite symmetric strip, since current is not flowing there (see Fig. 2(a-b)). The nucleation of new DWs into the R strip is possible after the driven DW splits at the opposite edge of the structure, depending on the applied pulse intensity $J$ ($J_1$ or $J_2$) and duration $d$. If $J$ is greater than a critical current ($J_C$), two new DWs are obtained: one that is simultaneously formed and driven by the same applied current along the R strip, and another one that \lq regenerates' the confined DW but now of opposite type (\lq \textit{down}-\textit{up}' in the sixth timestamp in Fig. 2(b)) and placed at the other symmetric strip. Nucleated DWs in the R strip will be driven independently of activating either $J_1$ or $J_2$. Thus, by alternatively applying current pulses between the TL or BL strip and the R strip, \textit{up} or \textit{down} domains can be simultaneously nucleated and driven.

\begin{figure}[t]
\centering 
\includegraphics[width=13cm]{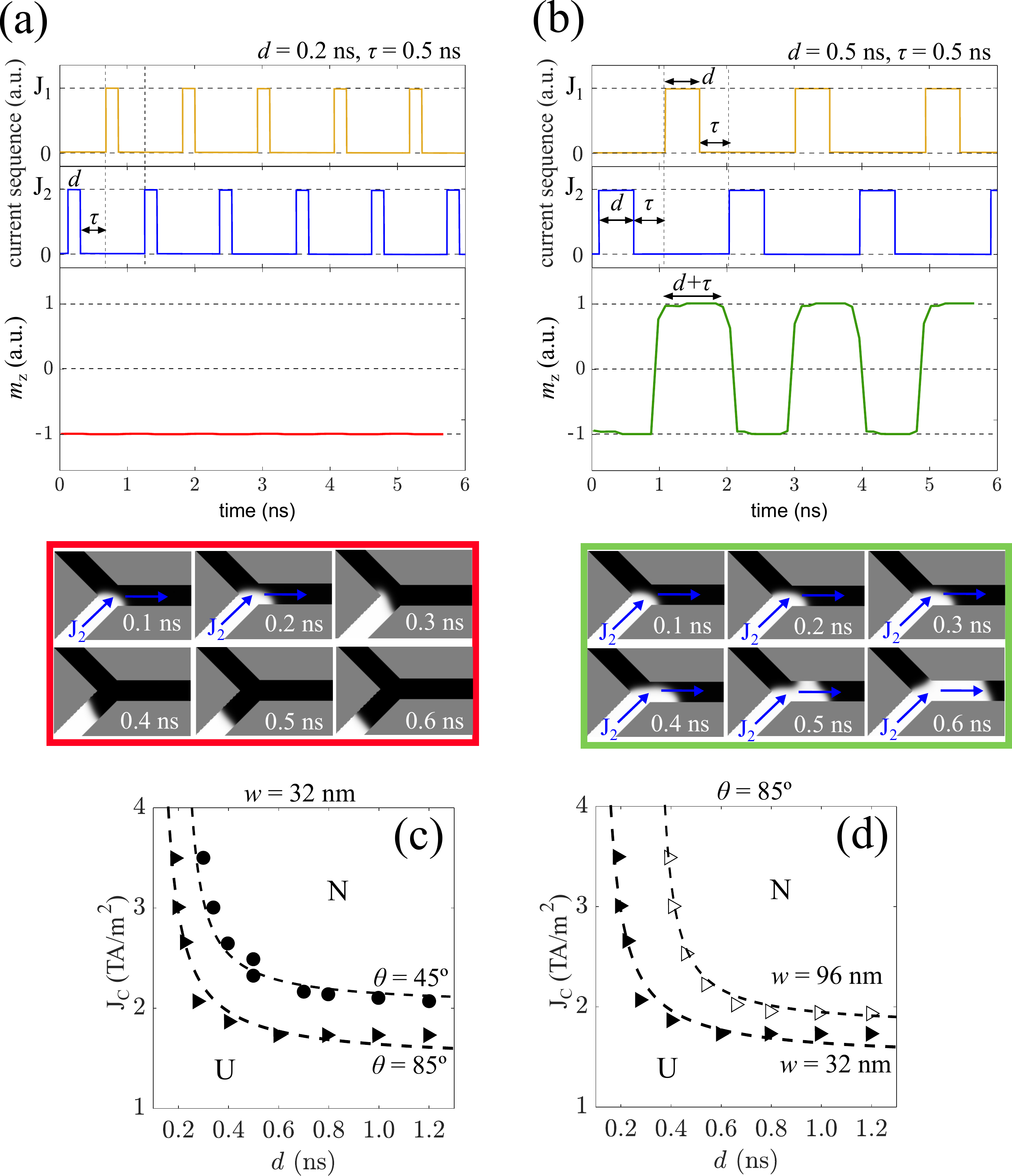}
\caption{(a)-(b) (Top panels) Current pulse sequences for two exemplary cases and (middle panels) their associated read-out signal (i.e. normalized $m_z$ at the \lq detector position' or orange area), as a response to the pulses. (Bottom panels) Snapshots of the instantaneous DW position for a single $J_2$ pulse excitation showing the out-of-plane magnetization, $m_z$, for the two states found in simulations. (c) Summarized full micromagnetic results for $J_C$ vs $d$ for $\theta=45$º (circles) and $\theta=85$º (triangles). Full symbols indicate the $J_C-d$ \lq boundary' beyond which \lq Nucleation' state was achieved. Dashed curves are for guiding the eye. (d) Summarized results for two different widths, 96 nm and 32 nm, and $\theta=85$º.}  \label{fig2}
\end{figure}

\par Then, regarding practical purposes, one question arises: how could we optimise information (domain) density in the R strip? To get further insight, Fig. 2(c) shows a summarized state/phase diagram for each combination of $J_C$ and $d$ explored in simulations for $\theta=45$º (circles) and $\theta = 85$º (triangles) and $w=32$ nm. Similar results were also obtained for other $\theta<90$º. Also, for $w=96$ nm or  $w=10$ nm. Fig. 2(d) shows similar qualitative results for $w=96$ nm. A more detailed analysis on the influence of $w$ is provided in the supplementary material. As expected, the critical current above which DW nucleation is achieved, reduces with increasing $d$. Also, the $J_C(d)$ curve is displaced downwards with reducing $w$ and/or increasing $\theta$. As discussed in the supplementary material, the efficiency of the SOT promoting the DW nucleation is enhanced as $\theta$ increases for a given $w$. Besides, one may expect that a higher domain density in the R strip can be obtained for high (small) $J$ and small (high) $d$, as the DW will be moving faster for a fixed time (or for longer time for a fixed velocity). However, at the cases where $J$ or/and $d$ are very high, undesirable effects arise due to spontaneous domain nucleation along the strip \cite{packed}, resulting in a faulty sequence, as observed in our simulations (not shown). Also, smaller $J_C$ require larger $d$ for DW nucleation, which leads to wider domains and therefore reduces the information density. Using wider strips would further allow to reduce $J_C$, but at the expense of increasing $d$. For all the above, an optimal density of domains is expected when both $J_C$ and $d$ are minimised, and by increasing $\theta$ and/or reducing $w$.

\begin{figure}[ht]
\centering 
\includegraphics[trim=0cm 0cm 0cm 0cm, clip=true, width=14cm]{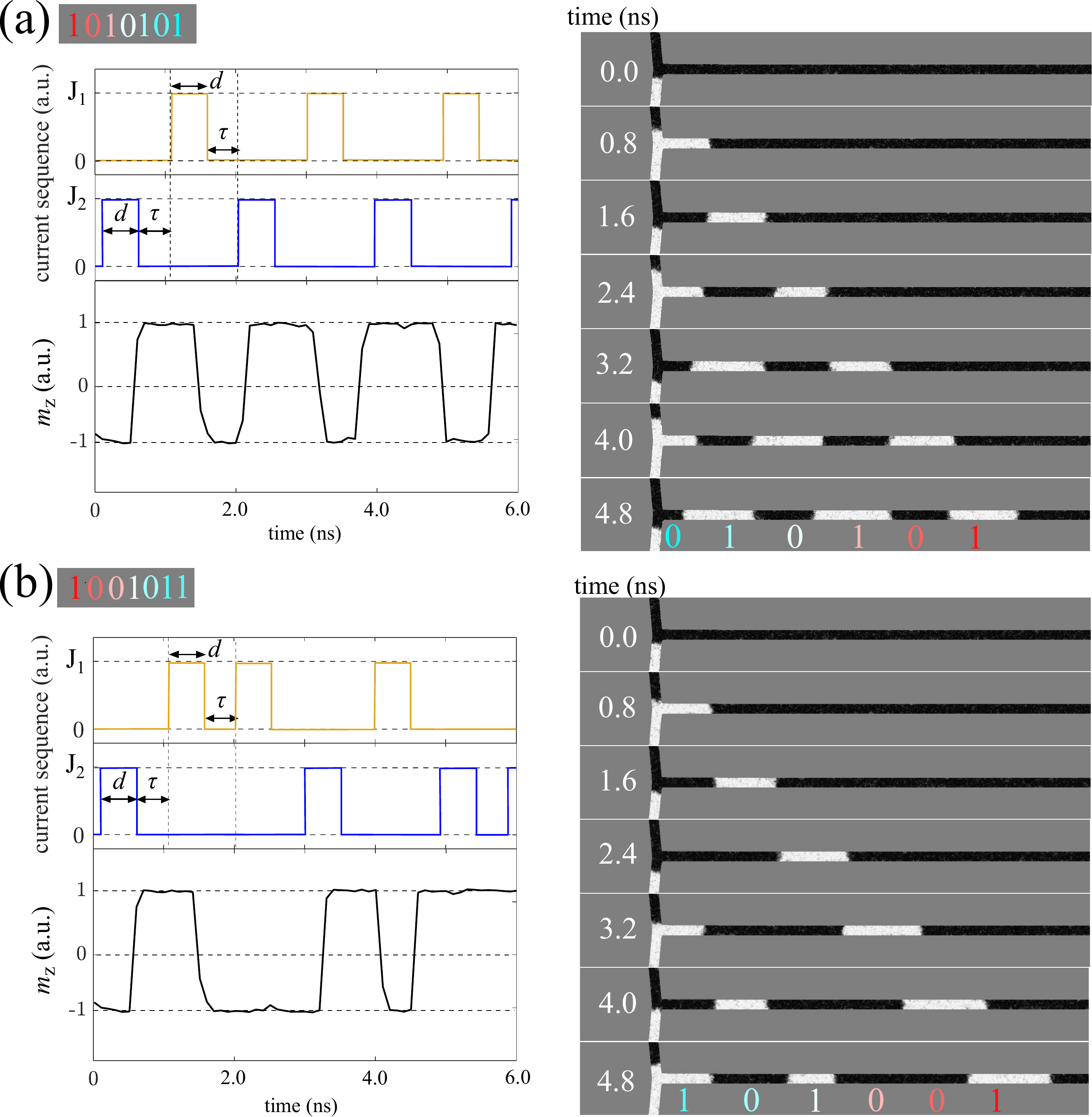}
\caption{ (a) Current pulse sequence (top left panel) and normalized $m_z$ at the detector area (bottom left panel) for $J=2.0$ TA/m$^2$ and $d=\tau=0.4$ ns for the bit sequence '1010101' ($J_1 \sim$ $'0'$ and $J_2 \sim$ $'1'$). Timestamps of $m_z$ are shown in the right panels (normalized, black is $-1$ and white is $+1$) under realistic conditions. (b) Equivalent results for the bit sequence '1001011'.}  \label{fig4}
\end{figure}

\par So far, we explored the performance of the proposed method under ideal conditions, i.e. no thermal fluctuations ($T= 0$ K), no grain structure nor rounded geometrical features. To get a better insight, we explore the performance of a more realistic strip accounting for thermal fluctuations at room temperature ($T = 300$ K), grains and rounded kinks, with $\theta=85$º. Following from previous studies \cite{angular_dependence}, the grains are modeled as a random deviation of maximum 10$\%$ of $K_U$ from its original direction ($\vec{u}_z$) in irregular sized regions of 3 nm radius on average. Rounded kinks of radius 30 nm are considered. 

\par  Two bit input sequences \lq 1-0-1-0-1-0-1' and \lq 1-0-0-1-0-1-1', encoded into $m_z$ by the pulse sequences $ J_2 \rightarrow J_1 \rightarrow J_2 \rightarrow J_1 \rightarrow J_2 \rightarrow J_1 \rightarrow J_2$ and $ J_2 \rightarrow J_1 \rightarrow J_1 \rightarrow J_2 \rightarrow J_1 \rightarrow J_2 \rightarrow J_2$ (shown in the top graphs in Fig. 3(a) and (b)) respectively, are simulated for $J= 2.0$ TA/m$^2$ and $d =\tau= 0.4$ ns under realistic conditions for a 32 nm wide strip. Fig. 3(a-b) shows $m_z$ as a function of time (bottom graphs), and timestamps of the nucleated \lq domain sequence' for both bit sequences. They reveal that the input current-pulse signals are effectively replicated along the R strip and correctly detected through $m_z$ under realistic conditions. Therefore, nucleation of domain sequences coherent with the input pulses is demonstrated as proof-of-concept. Additional simulations were also performed taking into account the influence of a Field-Like (FL-SOT) torque comparable in magnitude to the SL-SOT. The results, presented in the supplementary material, indicate that the working principle of our proposed method remains valid in the presence of FL-SOT. Moreover, the role of the non-uniform Joule heating has been also analyzed as described in the supplementary material. Such results point out that the proposed nucleation and shifting method still remains valid.

\par To conclude, we have devised a method to nucleate DWs with just three terminals, purely by current methods without requiring multilayer attachments, external magnetic fields or actively modifying the magnetic properties of the strip. A symmetric three-strip structure with a confined chiral Neel DW in the center is used as an optimal DW nucleator as a function of $J$ and $d$. Our design is simpler than those found in literature that are based on similar principles \cite{cross,doi:10.1063/1.4998216_selective}, and typically require multiple terminals for nucleating and driving DWs not concurrently \cite{doi:10.1063/1.4998216_selective} or locally attached metallic wires for injecting currents. Compared to recent proposals \cite{article_dao}, narrower domain (bit) widths of about 150 nm (three times narrower) in average are obtained for $J=2.0$ TA/m$^2$ and $d = 0.4$ ns, which gives an average greater bit density of $\sim$6 bits/$\mu$m in the R strip. Note that due to the long-range dependence of the Oersted field, a standard bit-line architecture is not useful to nucleate subsequent DWs without perturbing the information already coded in up or down domain between them (see details in Ref. \cite{doublestrip}). However, once the first DW is placed close to the junction with the R strip, our method does not require double \lq bit lines', external multilayer attachments, magnetic fields, nor manipulating magnetic anisotropy to nucleate and shift subsequent DWs along the \lq recording strip' in a controlled manner. Note that these techniques may be only used once to generate the confined DW, but are no longer required to sustain the process. Therefore, the proposed method significantly reduces the technological constraints for an iterative DW nucleation and the writing and transmitting of bits encoded in \textit{up} or \textit{down} domains.
\vspace{1cm} 

\textbf{Supplementary Material}

\par The provided supplementary material includes: (i) micromagnetic simulations of a conventional nucleation method for the initial DW, (ii) a detailed discussion on the reasons that may contribute to the reduction of $J_C$ when the aperture angle ($\theta $) increases, (iii) additional results for narrower strips ($w$ = 10 nm), (iv) the analysis of  the Field-Like (FL) torque, and (v) an investigation of the Joule heating effect on the proposed nucleation method.

\vspace{1cm} 
\textbf{Acknowledgments}
\par This work was supported by project SA114P20 from Junta de Castilla y León (JCyL), and partially by projects PID2020117024GB-C41 funded by Ministerio de Ciencia e Innovación, MAT2017-87072-C4-1-P from the Ministry of Economy, Spanish government, SA299P18 from JCyL, and MAGNEFI, from the European Commission, European Union. 

\vspace{1cm} 
\textbf{Data Availability}

The data that support the findings of this study are available from the corresponding author upon reasonable request.

\bibliography{library}

\end{document}